\documentclass[aps,preprint,groupedaddress,superscriptaddress,longbibliography]{revtex4-1}

\usepackage{graphicx}
\usepackage{dcolumn}
\usepackage{bm}
\usepackage{multirow}
\usepackage{color}
\usepackage{ulem}

\begin{document}

\title{Thickness-Dependent Interlayer Coupling and Semiconductor-to-Semimetal Crossover in Arsenene Multilayers}

\author{Jeonghwan Ahn}
\email{kindazet@gmail.com}
\affiliation{The Anthony J. Leggett Institute for Condensed Matter Theory, Department of Physics, University of Illinois at Urbana-Champaign, Urbana, IL 61801, USA
}
\affiliation{Materials Science and Technology Division, Oak Ridge National Laboratory, Oak Ridge, Tennessee 37831, USA}
\author{Seoung-Hun Kang}
\email{shkang@kisti.re.kr}
\affiliation{Research Center for Technology Commercialization, Korea Institute of Science and Technology Information (KISTI), Seoul, Korea}
\author{Jaron T. Krogel}
\email{krogeljt@ornl.gov}
\affiliation{Materials Science and Technology Division, Oak Ridge National Laboratory, Oak Ridge, Tennessee 37831, USA}

\date{\today}

\begin{abstract}
Interlayer interactions in layered materials are often assumed to transfer from the bilayer to the bulk, but this assumption can fail when chemically active out-of-plane orbitals participate in bonding. We combine diffusion quantum Monte Carlo (DMC) and density functional theory (DFT) to determine how interlayer coupling evolves in arsenene multilayers. DMC shows that bulk gray arsenic is compact, whereas the corresponding few-layer structures remain at substantially larger interlayer separations despite sharing the same nominal A$_{1}$B$_{-1}$ adjacent-layer registry. Registry alone therefore does not determine the bonding regime; thickness and coordination reshape the interlayer interaction. Among the tested functionals, SCAN+rVV10 most closely reproduces DMC equilibrium separations and stacking energetics. Using the DMC-benchmarked SCAN+rVV10 calculations, we predict a thickness-driven stacking sequence from A$_{1}$A$_{1}$ to A$_{1}$B$_{1}$ and finally bulk-like A$_{1}$B$_{-1}$. The structural crossover coincides with a stacking-dependent DFT band-gap collapse driven by enhanced interlayer As p$_{z}$ hybridization.

\end{abstract}

\maketitle

\section*{Introduction}
Interlayer interactions are a defining ingredient in layered materials. They determine not only the preferred stacking geometry and the equilibrium layer spacing, but also the low-energy electronic structure, the lattice response, and relative phase stability. This sensitivity becomes particularly important when the electronic states near the Fermi level have an appreciable out-of-plane character.  In such systems, even a modest change in registry or interlayer distance can alter orbital overlap between adjacent layers and drive a qualitatively different electronic response. A microscopic description of interlayer coupling is therefore essential for understanding how layered materials evolve from the few-layer limit to the bulk.

Arsenene is a useful example of this problem. Its buckled honeycomb lattice and lone-pair-derived out-of-plane states make the interlayer interaction highly sensitive to both local registry and layer separation~\cite{zhu2015strain,kecik2016stability,ersan2019two}. The resulting energetics are governed by a delicate balance between dispersive attraction, short-range repulsion, and direct interlayer hybridization. This balance is closely tied to the electronic structure: monolayer arsenene is predicted to be semiconducting~\cite{kamal2015arsenene,shah2020experimental}, while bulk gray arsenic is semimetallic~\cite{bullett1975density,gonze1990first,xu1993tight,zhao2017magnetotransport}. Arsenene therefore lies in a regime in which small errors in the treatment of interlayer coupling can change the predicted stacking energetics, equilibrium spacing, and the onset of low-energy metallic states.

Despite this importance, a consistent picture of interlayer interactions in arsenene has remained difficult to establish. Density functional theory (DFT) is a practical tool for exploring stacking- and thickness-dependent trends, but its predictions can be strongly dependent on the functional when dispersion, steric effects, and orbital hybridization occur on comparable energy scales. 
Previous high-level calculations have also led to different conclusions. A random phase approximation (RPA) study, which classified five symmetry-distinct bilayer stacking modes, identified A$_{1}$B$_{-1}$ as the lowest-energy bilayer structure~\cite{arcudia2020blue}. In contrast, diffusion quantum Monte Carlo (DMC) calculations favored A$_{1}$A$_{1}$ over A$_{1}$B$_{-1}$~\cite{kadioglu2018diffusion}. This apparent discrepancy is not yet definitive because previous comparisons did not place the competing bilayer stackings and bulk reference on the same many-body-referenced structural footing. 
It therefore remains unclear how the weakly bound few-layer regime connects to compact bulk gray arsenic, and whether a nominally identical adjacent-layer registry implies the same bonding regime across thickness.

The key advance of the present work is not simply the identification of a preferred bilayer stacking but the construction of a many-body-referenced link between the few-layer and bulk limits of arsenene. We combine DMC energy minimization for bilayer arsenene with a DMC benchmark against bulk gray arsenic, allowing weakly bound and compact interlayer regimes to be compared within the same material. This comparison shows that bilayer A$_{1}$B$_{-1}$ remains in a large-spacing, weak-coupling regime, whereas the corresponding adjacent-layer registry in bulk gray arsenic adopts a much shorter spacing with covalent-like interlayer bonding character. The same nominal stacking label therefore does not uniquely define the interlayer interaction regime. Instead, local registry and thickness-dependent coordination act together to determine whether arsenene remains weakly coupled or develops compact interlayer bonding.

Among the density functionals tested here, SCAN+rVV10 most consistently reproduces the DMC interlayer separations and relative stacking energetics for the benchmarked systems. Using this DMC-benchmarked functional, we then follow the structural and electronic evolution over a wider range of thicknesses. Within the stacking families considered, arsenene evolves from A$_{1}$A$_{1}$ to A$_{1}$B$_{1}$ and finally to bulk-like A$_{1}$B$_{-1}$ with increasing layer number, rather than approaching the bulk as a simple continuation of the bilayer ground state. This structural crossover is accompanied by a stacking-dependent collapse of the DFT band gap, associated with enhanced interlayer As $p_{z}$ hybridization. Our results provide a microscopic picture of cooperative interlayer coupling in arsenene multilayers and suggest that orbitals with chemically active out-of-plane character can give rise to thickness-dependent bonding regimes that are not apparent from either the bilayer or bulk limit alone.

\section*{Results}
\label{sec:results}

\begin{figure*}
            \centering
            \includegraphics[width=6.0in]{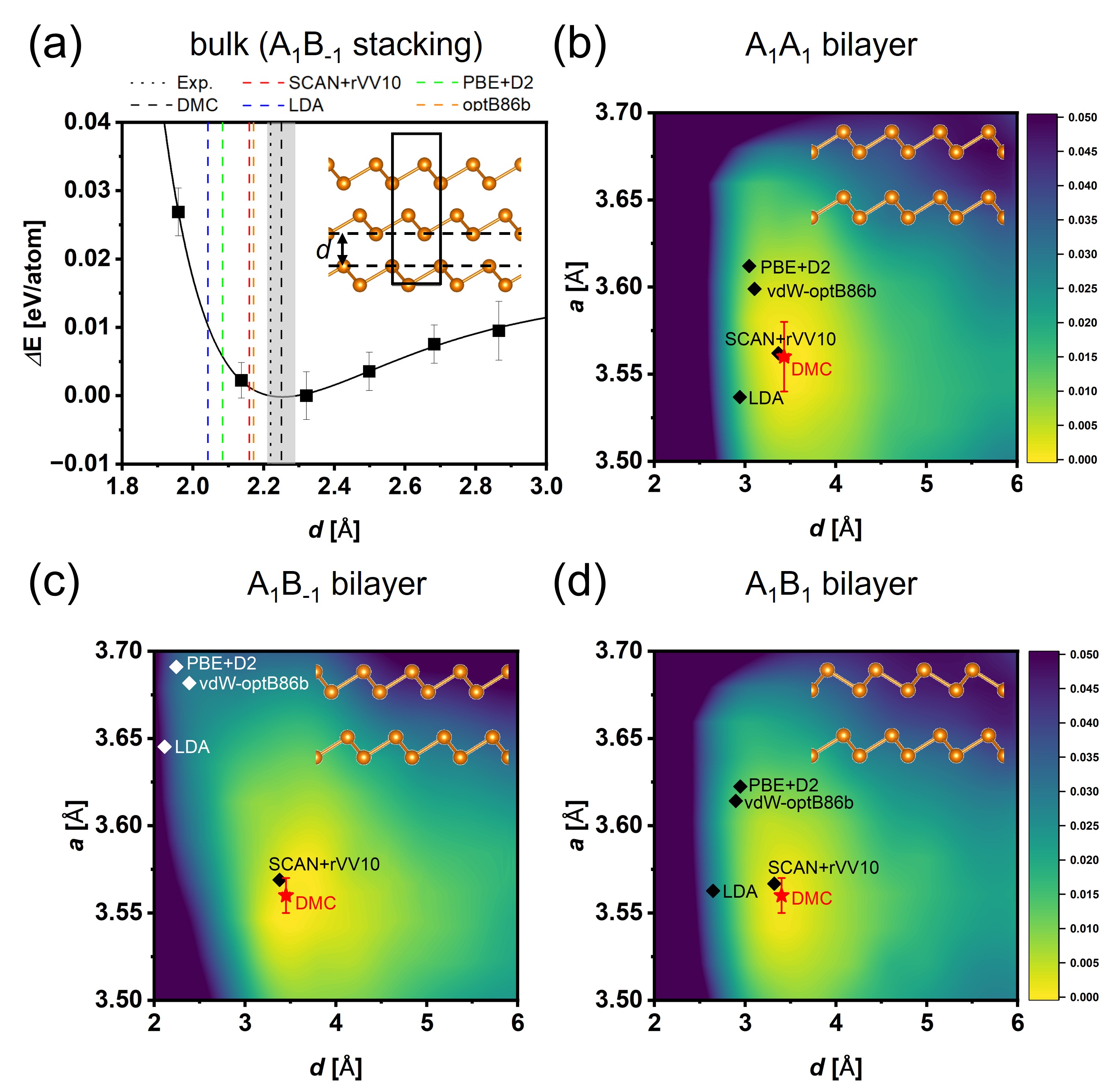}
            \caption{DMC benchmark of bulk gray arsenic and bilayer arsenene. (a) Relative total energy of bulk gray arsenic as a function of interlayer distance $d$ for the ABC stacking sequence, whose adjacent-layer registry corresponds to A$_1$B$_{-1}$. Vertical dashed lines indicate the equilibrium interlayer distances from experiment, DMC, SCAN+rVV10, LDA, PBE+D2, and vdW-optB86b. The inset shows the bulk structure and the definition of $d$. (b)--(d) Two-dimensional energy landscapes of bilayer arsenene as functions of interlayer distance $d$ and in-plane lattice constant $a$ for the A$_1$A$_1$, A$_1$B$_{-1}$, and A$_1$B$_1$ stackings, respectively. Symbols indicate the minima predicted by DMC and the different DFT methods, and the insets show representative structures for each bilayer stacking. The color scale in panels (b)–(d) gives the relative energy per atom in units of eV/atom.} 
            \label{fig:DMC_benchmark_distance}
\end{figure*}

We begin by benchmarking the DMC description against bulk gray arsenic, for which experimental structural data are available. Bulk gray arsenic adopts an ABC stacking sequence, whose adjacent-layer registry corresponds to A$_1$B$_{-1}$ in the notation of Fig.~\ref{fig:DMC_benchmark_distance}(a). As shown in Fig.~\ref{fig:DMC_benchmark_distance}(a), the total DMC energy exhibits a minimum at an interlayer distance of 2.25(3)~\AA, obtained using a $3 \times 3 \times 2$ supercell, in good agreement with the experimental value of 2.21--2.24~\AA~\cite{schiferl1969crystal}. The DFT equilibrium distances span a wide range and lie on the shorter-distance side of the DMC and the experimental values, with SCAN+rVV10 and vdW-optB86b remaining closer than LDA and PBE+D2. In the experimental gray-arsenic structure, the closest interlayer As--As distance is about 3.12~\AA, only moderately longer than the intralayer As--As bond length of approximately 2.52~\AA~\cite{schiferl1969crystal}. This compact geometry, together with the charge redistribution discussed in the following, is consistent with covalent-like interlayer bonding in bulk gray arsenic. The agreement with experiment establishes DMC as a reliable reference for interlayer energetics in arsenene multilayers.

\begin{table}[t]
\centering
\caption{
Equilibrium interlayer distance \(d\) and in-plane lattice constant \(a\) of bilayer arsenene for the three stacking geometries shown in Fig.~1.
}
\label{tab:fig1-minima}
\normalsize
\setlength{\tabcolsep}{7pt}
\renewcommand{\arraystretch}{1.20}
\begin{tabular}{l|cc|cc|cc}
\hline\hline
& \multicolumn{2}{c|}{\(A_1A_1\text{~stacking}\)}
& \multicolumn{2}{c|}{\(A_1B_{-1}\text{~stacking}\)}
& \multicolumn{2}{c}{\(A_1B_1\text{~stacking}\)} \\
\cline{2-7}
&
\(d\text{~[\AA]}\) & \(a\text{~[\AA]}\)
& \(d\text{~[\AA]}\) & \(a\text{~[\AA]}\)
& \(d\text{~[\AA]}\) & \(a\text{~[\AA]}\) \\
\hline
DMC
& \(3.43(5)\) & \(3.56(2)\)
& \(3.40(2)\) & \(3.56(1)\)
& \(3.45(4)\) & \(3.56(1)\) \\
LDA
& 2.94 & 3.54
& 2.11 & 3.65
& 2.65 & 3.56 \\
SCAN+rVV10
& 3.37 & 3.56
& 3.38 & 3.57
& 3.32 & 3.57 \\
PBE+D2
& 3.05 & 3.61
& 2.24 & 3.69
& 2.95 & 3.62 \\
vdW-optB86b
& 3.10 & 3.60
& 2.39 & 3.68
& 2.90 & 3.61 \\
\hline\hline
\end{tabular}
\end{table}

The bilayer energy landscapes in Figs.~\ref{fig:DMC_benchmark_distance}(b)--\ref{fig:DMC_benchmark_distance}(d) reveal a qualitatively different picture from the bulk. For all three bilayer stackings, A$_1$A$_1$, A$_1$B$_{-1}$ (bulk-like), and A$_1$B$_1$, the DMC minima occur at substantially larger interlayer separations than in the bulk, clustered near $d \sim 3.4$--$3.5$~\AA\ and $a \sim 3.55$~\AA~ (see Table~\ref{tab:fig1-minima}). The DFT minima are more dispersed and generally shifted toward shorter interlayer distances. This discrepancy is most pronounced for bilayer A$_1$B$_{-1}$, where LDA, PBE+D2, and vdW-optB86b favor markedly compressed geometries relative to DMC, while SCAN+rVV10 remains closest to the DMC minimum. A similar trend is seen for A$_1$A$_1$ and A$_1$B$_1$, where SCAN+rVV10 again tracks the DMC minimum more closely than the other functionals.

The contour plots also show that the short-range repulsive part of the interlayer energy landscape is strongly stacking dependent. Among the three bilayer stackings, A$_1$B$_{-1}$ maintains relatively low energies down to smaller interlayer distances than A$_1$A$_1$ or A$_1$B$_1$, indicating that compression is accommodated more readily in this registry. This feature is consistent with the bulk result in Fig.~\ref{fig:DMC_benchmark_distance}(a), where the ABC-stacked structure, whose adjacent-layer registry corresponds to A$_1$B$_{-1}$, is stabilized at much shorter interlayer distance. Figures~\ref{fig:DMC_benchmark_distance}(b)--\ref{fig:DMC_benchmark_distance}(d) therefore show not only that the bilayer minima lie in a markedly different interlayer-distance regime from the bulk, but also that the energetic cost of compression depends sensitively on stacking. 

\begin{figure*}
            \centering
            \includegraphics[width=6.0in]{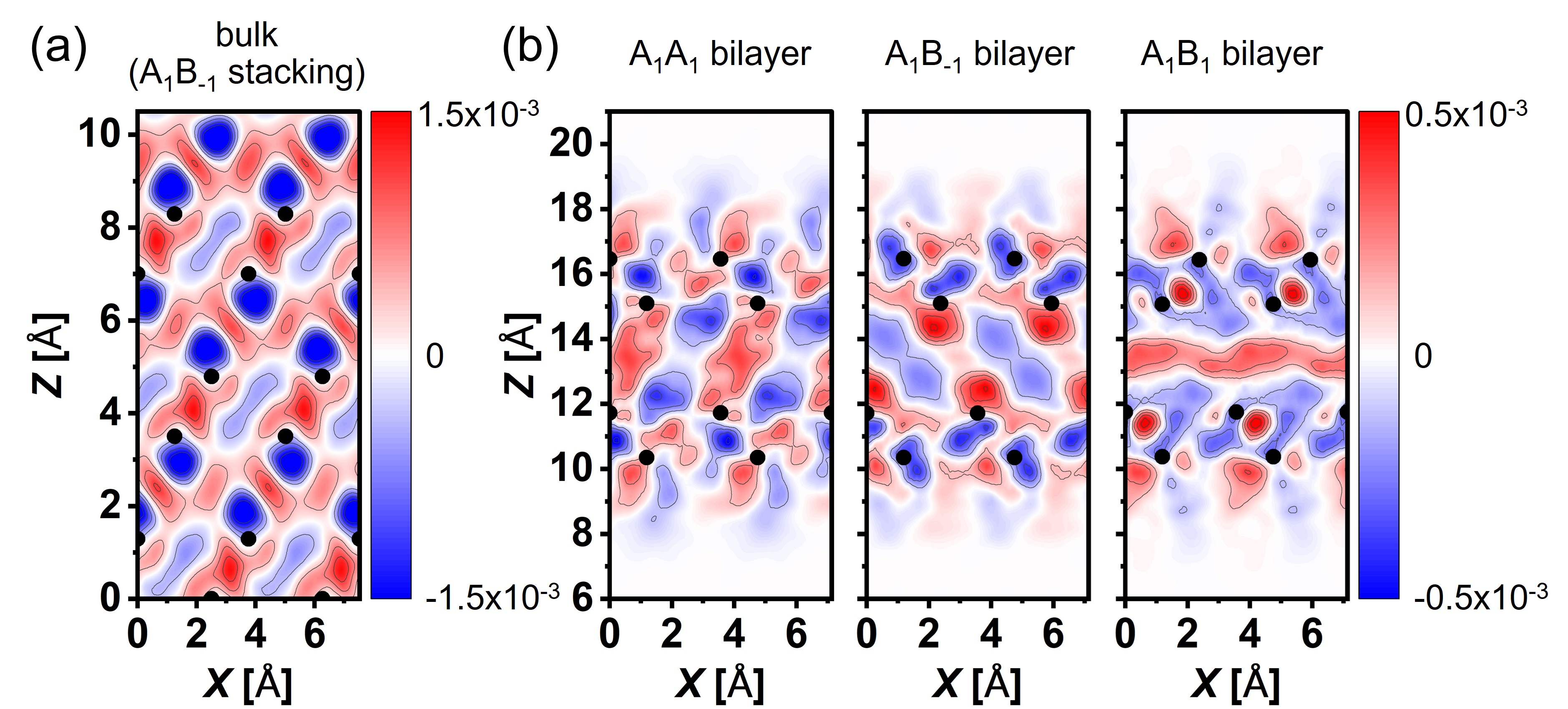}
            \caption{DMC charge-density redistribution associated with interlayer binding in bulk and bilayer arsenene. (a) Bulk A$_1$B$_{-1}$ stacking, evaluated as $\Delta \rho = \rho_{\mathrm{bulk}}-\rho_{\mathrm{upper}}-\rho_{\mathrm{middle}}-\rho_{\mathrm{lower}}$. (b) Bilayer A$_1$A$_1$, A$_1$B$_{-1}$, and A$_1$B$_1$, evaluated as $\Delta \rho = \rho_{\mathrm{bilayer}}-\rho_{\mathrm{upper}}-\rho_{\mathrm{lower}}$. Red and blue denote electron accumulation and depletion, respectively, and black circles mark the projected positions of As atoms. Units are \AA$^{-2}$.} 
            \label{fig:charge_density}
\end{figure*}

Figure~\ref{fig:charge_density} shows the charge-density difference associated with interlayer binding in bulk gray arsenic and bilayer arsenene. To make the weaker bilayer response visible, the bilayer panels are plotted with a color scale that is one third of that used for the bulk panel. In bulk A$_1$B$_{-1}$ displayed in Fig.~\ref{fig:charge_density}(a), electron accumulation and depletion form an ordered pattern across adjacent interlayer regions, with positive redistribution extending along interlayer contact directions rather than remaining confined to isolated atomic sites.
This extended redistribution is consistent with covalent-like interlayer bonding in the compact bulk structure. The bilayers show a much weaker and localized response. Even bilayer A$_1$B$_{-1}$, which shares the same nominal adjacent-layer registry as bulk gray arsenic, does not reproduce the bulk-like redistribution pattern. 
Its charge response remains local and interrupted, as in the other bilayer stackings, rather than developing into an extended interlayer network. Therefore, these maps distinguish two limiting regimes of interlayer coupling in arsenene: a localized weak-coupling response in the bilayer limit and a developed interlayer-bonding pattern in the compact bulk limit, even for the same nominal A$_1$B$_{-1}$ registry.

The bilayer maps further show that the weak bilayer response is not purely dispersive. In A$_1$B$_{-1}$ and A$_1$A$_1$, electron accumulation appears along the closest interlayer contact directions, accompanied by depletion in nearby interlayer regions. This contact-centered redistribution suggests a finite orbital-overlap component to the interlayer coupling, consistent with interlayer As $p_z$ hybridization. A$_1$B$_1$ displays a distinct pattern, in which electron accumulation spreads more broadly through the middle of the interlayer region, with less pronounced depletion at the central interface. Compared with contact-centered patterns in A$_1$A$_1$ and A$_1$B$_{-1}$, this broader redistribution suggests that A$_1$B$_1$ may support more spatially distributed interlayer As $p_z$ orbital-overlap pathways as additional layers are added. 
The real-space charge response therefore provides a microscopic link between the stacking-dependent interlayer interaction shown in Fig.~\ref{fig:charge_density} and the stronger reconstruction of the low-energy electronic structure of A$_1$B$_1$ near the gap edges and, at larger thickness, near the Fermi level, compared to the other stacking sequences.
Thus, even in the bilayer limit, the interlayer interaction contains an electronic component beyond a purely van der Waals picture, while the bulk structure represents a fully developed interlayer-bonding regime.

\begin{figure*}
            \centering
            \includegraphics[width=6.5in]{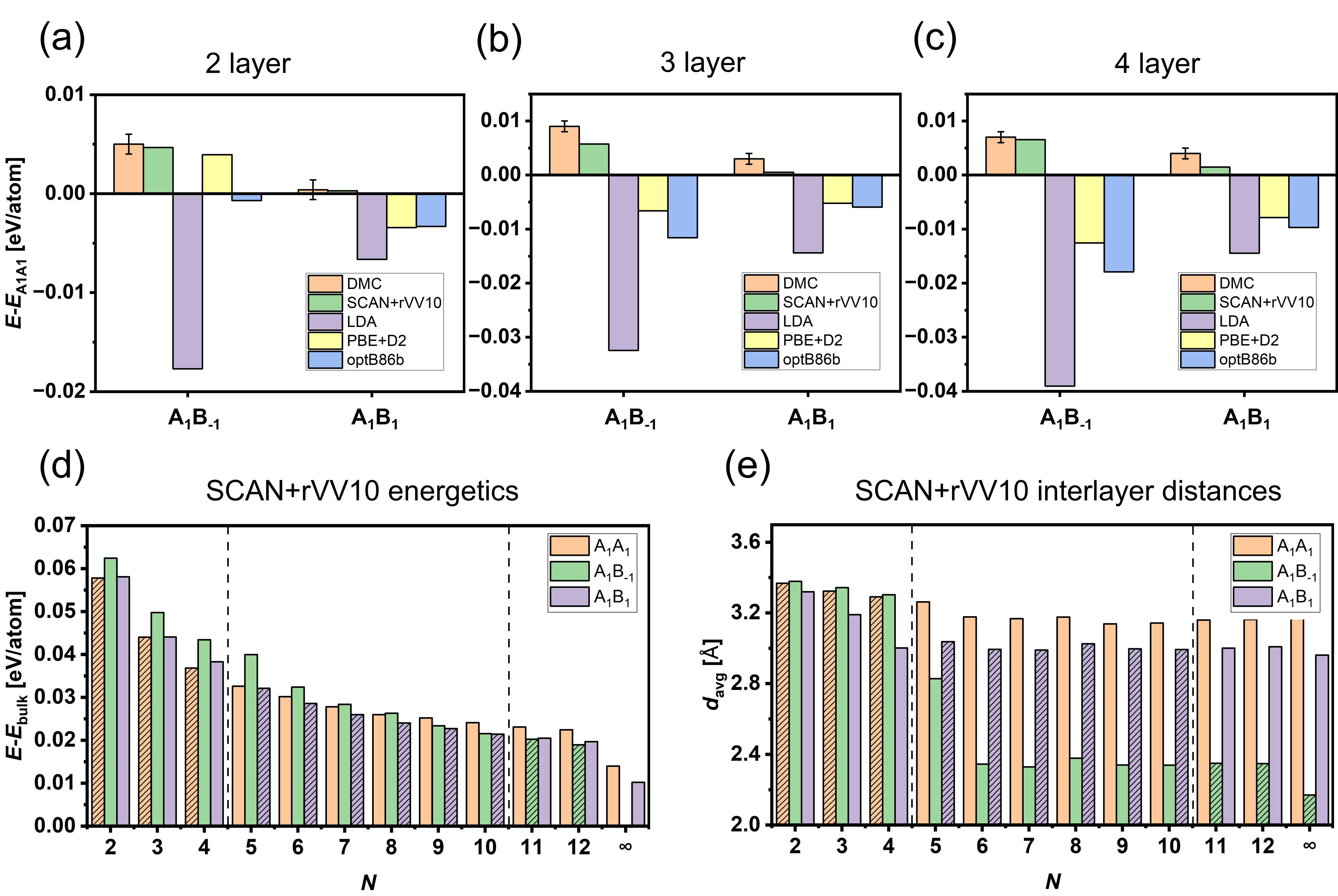}
            \caption{DMC benchmark and SCAN+rVV10 thickness evolution of stacking energetics. (a)--(c) Relative energies of the A$_1$B$_{-1}$ and A$_1$B$_1$ stackings with respect to A$_1$A$_1$ for 2-, 3-, and 4-layer arsenene, comparing DMC with SCAN+rVV10, LDA, PBE+D2, and vdW-optB86b. (d) SCAN+rVV10 energies of the A$_1$A$_1$, A$_1$B$_{-1}$, and A$_1$B$_1$ stackings as a function of layer number $N$, referenced to bulk gray arsenic. (e) Average interlayer distance $d_{\mathrm{avg}}$ of the three stackings as a function of $N$ from SCAN+rVV10. In panels (d) and (e), hatching marks the bar corresponding to the lowest-energy stacking at each layer number. The vertical dashed lines indicate the stacking transitions that separate the A$_1$A$_1$, A$_1$B$_{1}$, and A$_1$B$_{-1}$ stability regions.}
             \label{fig:relative_energy}
\end{figure*}

Figure~\ref{fig:relative_energy}(a)--\ref{fig:relative_energy}(c) compares the DMC and DFT relative energies of the three low-energy stacking families for 2-, 3-, and 4-layer arsenene. 
In all three thicknesses, DMC identifies A$_1$A$_1$ as the lowest-energy stacking among the structures considered, while A$_1$B$_1$ remains a nearby competitor, and A$_1$B$_{-1}$ remains higher throughout this range of few layers. A near degeneracy between A$_1$A$_1$ and A$_1$B$_1$ appears only in the bilayer. This ordering provides a direct many-body benchmark for the low-thickness regime and shows that the bulk-like A$_{1}$B$_{-1}$ registry is not immediately favored when the system is reduced to a few layers.

The comparison with DFT shows that the predicted stacking hierarchy is sensitive to the functional. LDA, PBE+D2 and vdW-optB86b tend to over-stabilize the most compressed stackings, most notably A$_{1}$B$_{-1}$. This trend is consistent with the bilayer energy landscapes in Fig.~\ref{fig:DMC_benchmark_distance}, where these functionals also favor shorter interlayer separations than DMC. By contrast, SCAN+rVV10 most consistently reproduces the DMC ordering for 2-, 3-, and 4-layer arsenene. We therefore use SCAN+rVV10 as a DMC-benchmarked framework to examine the thickness evolution beyond the range directly accessible to DMC.

Having established this agreement, we use SCAN+rVV10 to follow the thickness evolution over a wider range. Figure~\ref{fig:relative_energy}(d) shows that the lowest-energy stacking among the considered families changes sequentially with the number of layers. The hatched bars mark the lowest-energy stacking in each $N$: A$_1$A$_1$ is favored for $N \le 4$, A$_1$B$_1$ becomes lowest for $5 \le N \le 10$, and bulk-like A$_1$B$_{-1}$ takes over for $N \ge 11$. The bulk approach is therefore not a direct continuation of the bilayer ground state. 
Arsenene first stabilizes in the A$_{1}$A$_{1}$ few-layer regime, then passes through A$_{1}$B$_{1}$ before the compact A$_{1}$B$_{-1}$ structure becomes energetically preferred. This sequence indicates that the few-layer energy landscape cannot be reduced to a simple thin-film continuation of the bulk stacking picture.

The same thickness evolution is not obtained with vdW-optB86b and PBE+D2. In both functionals, $A_1B_1$ is favored only at $N=2$, while $A_1B_{-1}$ becomes the lowest-energy stacking already at $N=3$ and remains favored through the bulk limit (see Supplementary Fig. 1). These results give a different thickness-dependent energy profile from the DMC-benchmarked SCAN+rVV10 trend. Instead of the stepwise $A_1A_1 \rightarrow A_1B_1 \rightarrow A_1B_{-1}$ sequence, vdW-optB86b and PBE+D2 predict early stabilization of bulk stacking $A_1B_{-1}$. This comparison further shows that the predicted route from the few-layer limit to the bulk is sensitive to the treatment of interlayer energetics.

Figure~\ref{fig:relative_energy}(e) clarifies the microscopic origin of this sequence. The average interlayer distance of A$_1$A$_1$ remains large and changes only weakly with thickness, while A$_1$B$_1$ stabilizes at an intermediate spacing near 3.0~\AA, placing it between the weakly coupled few-layer regime and the compact bulk-like structure. By contrast, A$_1$B$_{-1}$ shows the strongest thickness dependence. It starts in a large-spacing regime for small $N$, then undergoes a rapid reduction in d$_{\text{avg}}$ and eventually approaches the compact bulk-like limit. 
In other words, A$_1$B$_{-1}$ is the only stacking among those examined that can fully exploit thickness to access a compact interlayer geometry. 
This is consistent with the contour plots in Fig.~\ref{fig:DMC_benchmark_distance}, where A$_1$B$_{-1}$ accommodates shorter interlayer distances at lower energetic cost than A$_1$A$_1$ or A$_1$B$_1$.
The PBE+D2 and vdW-optB86b results in Supplementary Fig.~1 provide the contrasting limit: their A$_1$B$_{-1}$ structures remain close to the compact bulk-like interlayer distance from $N=2$ through the bulk limit, rather than showing a delayed thickness-driven compression obtained with DMC-benchmarked SCAN+rVV10 calculations. This persistent compactness is consistent with their energy profiles, where A$_1$B$_{-1}$ already becomes the lowest-energy stacking at $N=3$.
Together, these results show that the stacking crossover is not governed by the registry alone, but by the thickness-dependent ability of each stacking sequence to develop compact interlayer bonding.

\begin{figure*}
            \centering
            \includegraphics[width=6.5in]{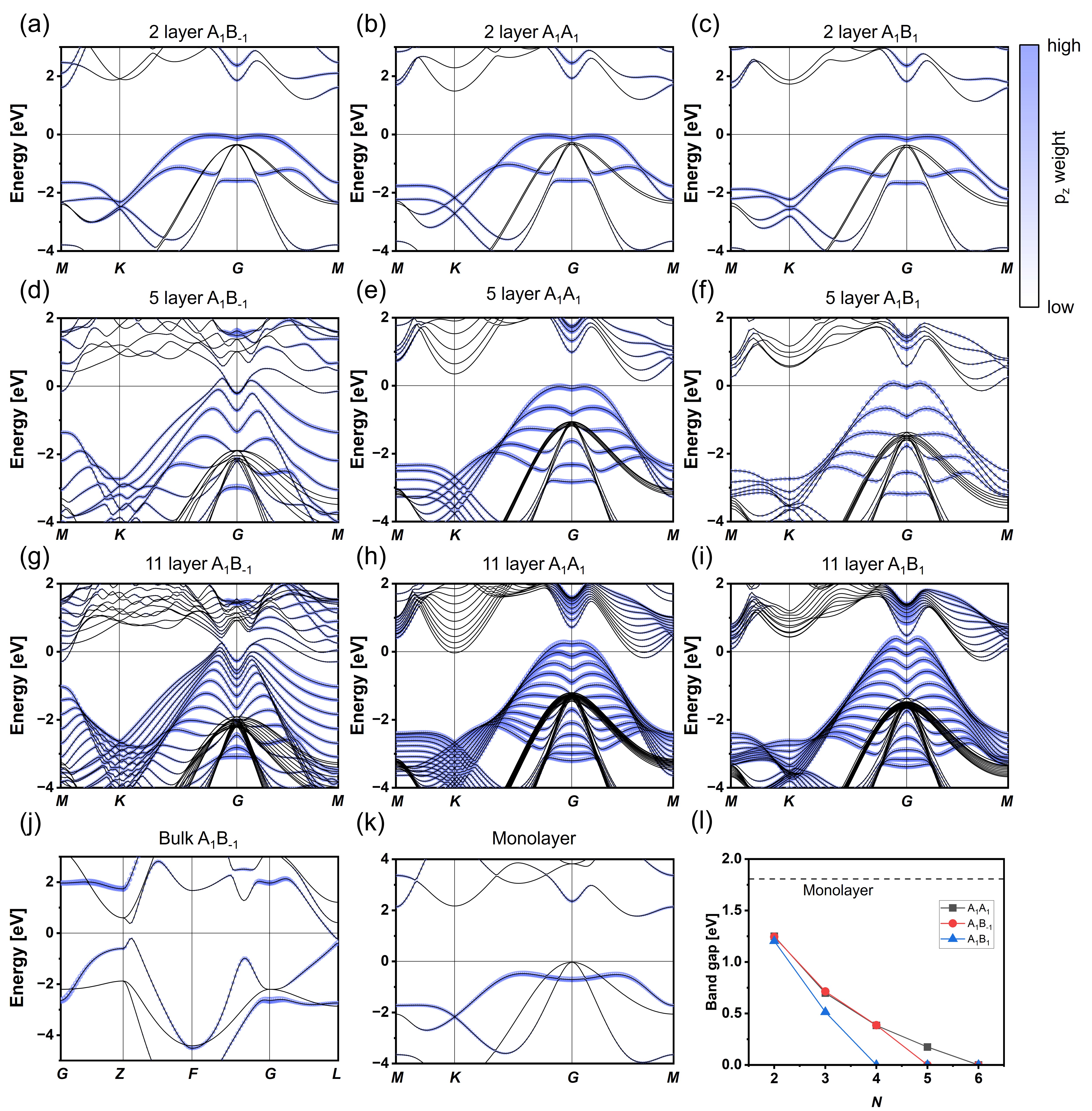}
            \caption{Orbital-projected band structures colored by As $p_z$ weight for (a) 2-layer A$_1$B$_{-1}$, (b) 2-layer A$_1$A$_1$, (c) 2-layer A$_1$B$_1$, (d) 5-layer A$_1$B$_{-1}$, (e) 5-layer A$_1$A$_1$, (f) 5-layer A$_1$B$_1$, (g) 11-layer A$_1$B$_{-1}$, (h) 11-layer A$_1$A$_1$, (i) 11-layer A$_1$B$_1$, (j) bulk A$_1$B$_{-1}$, and (k) monolayer arsenene. Blue intensity denotes the As-$p_z$ weight. (l) DFT-SCAN+rVV10 band gap as a function of layer number for the A$_1$A$_1$, A$_1$B$_{-1}$, and A$_1$B$_1$ stackings.} 
            \label{fig:band}
\end{figure*}

The structural crossover described above is accompanied by a stacking-dependent reconstruction of the low-energy electronic structure. The SCAN+rVV10 orbital-projected band structures in Fig.~\ref{fig:band} show that the low-energy states of arsenene multilayers already have substantial As $p_z$ character in the bilayer limit, but their thickness evolution is strongly dependent on stacking. The monolayer band structure in Fig.~\ref{fig:band}(k) provides the isolated-layer reference, where arsenene remains semiconducting before interlayer hybridization is introduced. In the 2-layer structures, all three stackings remain semiconducting and have appreciable As p$_{z}$ character near the gap edges. The bilayer bands therefore indicate that out-of-plane orbitals already participate in the low-energy electronic structure, although the stacking dependence remains modest at this thickness.

This difference becomes more pronounced at intermediate thickness. By 5 layers, A$_1$B$_1$ is metallic at the DFT level and A$_1$B$_{-1}$ is close to gap closure, while A$_1$A$_1$ remains semiconducting. The states approaching the Fermi level carry strong As p$_{z}$ weight, indicating that the gap reduction is not a rigid narrowing of the spectrum, but a stacking-selective enhancement of interlayer hybridization involving out-of-plane orbitals. The gap trends in Fig.~\ref{fig:band}(l) make this sequence explicit: the SCAN+rVV10 band gap decreases with thickness for all three stacking sequences, but the collapse occurs faster for A$_1$B$_1$, followed by A$_1$B$_{-1}$ and most gradually for A$_1$A$_1$.

Together with the interlayer distances in Fig.~\ref{fig:relative_energy}(e), these band structures connect the electronic crossover to the structural evolution. For A$_1$B$_1$, the gap closes as the structure reaches an intermediate spacing near 3.0~\AA, after which d$_{\text{avg}}$ changes only slightly with thickness. For A$_1$B$_{-1}$, the rapid reduction in $d_{\mathrm{avg}}$ coincides with the approach to metallization and is followed by a compact, bulk-like spacing in larger $N$. The metallic regime is therefore not accompanied by continued strong compression. Instead, once a stacking sequence develops sufficiently strong As p$_{z}$-mediated interlayer hybridization, its preferred interlayer spacing is already close to a saturated value. In this sense, metallization and flattening of $d_{\mathrm{avg}}$ are linked consequences of the same strengthening of the interlayer interaction.

By 11 layers, all three stackings exhibit developed $p_z$-derived manifolds near the Fermi level, but the resulting metallic states are not equivalent. In A$_1$B$_1$, strong $p_z$ weight remains concentrated in several bands near the Fermi level, consistent with a metallic state reached through the rapid upward shift of out-of-plane states. In contrast, 11-layer A$_1$B$_{-1}$ shows a broader redistribution of $p_z$ character over occupied and unoccupied branches. As shown in the bulk A$_1$B$_{-1}$ electronic structure in Fig.~\ref{fig:band}(j), 
a characteristic pair of p$_{z}$-weighted branches appears near $\Gamma$, with one branch below and the other above the Fermi level. The 11-layer A$_{1}$B$_{-1}$ band structure develops the same occupied-unoccupied p$_{z}$-weighted feature near the Fermi level. This correspondence indicates that the compact 11-layer A$_{1}$B$_{-1}$ geometry already carries a bulk-like reorganization of the low-energy electronic structure. Figure~\ref{fig:band} therefore connects the structural evolution in Fig.~\ref{fig:relative_energy} to a stacking-dependent electronic transition. Thickness alone does not determine when the gap closes or what kind of metallic state emerges. Instead, these electronic outcomes are determined by how each stacking reorganizes interlayer hybridization among As p$_{z}$-derived states.

\section*{Discussion}
\label{sec:discussion}

High-level studies have already established bilayer arsenene as a stringent test case for describing interlayer interactions. The previous DMC study provided an important many-body benchmark and supported A$_1$A$_1$ over A$_1$B$_{-1}$ for buckled bilayer arsenene, but it was restricted to two bilayer stackings~\cite{kadioglu2018diffusion}. In contrast, the RPA study broadened the structural landscape by examining all five symmetry-distinct buckled bilayer stackings and identified A$_1$B$_{-1}$ as the lowest-energy form of gray arsenene, with bilayer geometries optimized at the PBE+MBD level~\cite{arcudia2020blue}. These studies clarified different aspects of the problem, but did not place the relevant bilayer stackings and the bulk reference on the same many-body-grounded structural footing. The present work advances this picture by combining DMC-level bilayer geometry optimization with a DMC benchmark against bulk gray arsenic. This comparison shows that the apparent difference between the earlier DMC and RPA results is not simply a conflict in predicted stacking order. Instead, it reflects the fact that the same nominal A$_1$B$_{-1}$ registry can belong to distinct bonding regimes as arsenene evolves from the weakly coupled few-layer limit to compact bulk gray arsenic.

The main result of this work is that interlayer interactions in arsenene cannot be described by a single thickness-independent picture. DMC energy minimization in the structural subspace spanned by the in-plane lattice constant and interlayer separation identifies a weakly bound large-spacing regime in bilayer arsenene, whereas the DMC benchmark for bulk gray arsenic recovers the compact experimental interlayer spacing of the ABC-stacked structure. This contrast shows that the registry alone does not determine the character of interlayer coupling. Rather, the local registry and the thickness-dependent coordination environment jointly determine whether the system remains weakly coupled or develops compact covalent-like interlayer bonding.
This distinction also explains why approximate density functionals can diverge strongly in this material. Functionals that overstabilize compressed bilayer geometries also distort the energetic ordering among the competing few-layer stackings, whereas SCAN+rVV10 most consistently reproduces the DMC equilibrium separations and relative stacking energetics for the systems directly benchmarked here. 

Using this DMC-benchmarked functional framework, we predict a stepwise thickness evolution of the lowest-energy stacking among the considered families: A$_{1}$A$_{1}$ is favored in the few-layer limit, A$_{1}$B$_{1}$ becomes stable at intermediate thickness, and bulk-like A$_{1}$B$_{-1}$ is recovered only at a larger layer number. This sequence indicates that arsenene does not approach the bulk as a simple thin-film continuation of the gray-arsenic stacking picture. Instead, it passes through an intermediate A$_1$B$_1$ regime where moderate interlayer compression and enhanced As $p_z$-derived hybridization together stabilize a distinct multilayer structure before the compact A$_1$B$_{-1}$ registry becomes favorable. The corresponding electronic structures exhibit a stacking-dependent collapse of the SCAN+rVV10 band gap, with A$_1$B$_1$ reaching the gap closure more rapidly and A$_1$B$_{-1}$ eventually developing the most bulk-like metallic electronic structure. These results identify cooperative interlayer coupling, rather than layer number alone, as the microscopic origin of the structural and electronic crossover in arsenene multilayers.

More broadly, our results show that layered materials with chemically active out-of-plane states can host thickness-dependent bonding regimes that are not apparent from either limiting description alone: the weakly coupled bilayer or the compact bulk. In arsenene, the local registry controls how adjacent layers interact, while increasing thickness changes the coordination environment and enables different stackings to develop interlayer hybridization in distinct ways. This provides a route for tuning structural and electronic crossovers through stacking and thickness and establishes arsenene multilayers as a platform in which interlayer bonding, orbital hybridization, and metallization are cooperatively controlled.

\section*{Methods}
\label{sec:methodology}
Our fixed-node DMC simulations were carried out using the QMCPACK package~\cite{kim18,kent20}. As trial wave functions, we employed a Slater–Jastrow ansatz, composed of products of Slater determinants and Jastrow correlation factors, with terms up to three-body order to describe electron–electron–ion correlations. The Slater determinants were constructed from Kohn–Sham orbitals obtained in DFT-PBE calculations performed with QUANTUM ESPRESSO~\cite{giannozzi09}, using a plane-wave cutoff of 200 Ry and an $18 \times 18 \times 1$ ($18 \times 18 \times 4$) Monkhorst–Pack $k$-point mesh~\cite{monkhorst76} for few-layer arsenene (bulk gray arsenic). We used a norm-conserving pseudopotential developed within the ccECP scheme~\cite{wang2019new} for the As atom. To mitigate one-body finite-size effects, we adopted twist-averaged boundary conditions~\cite{lin01}, employing 16, 9, and 9 twist angles for supercells of size $3 \times 3 \times 1$, $4 \times 4 \times 1$, and $5 \times 5 \times 1$, respectively. The resulting twist-averaged DMC energies were then linearly extrapolated to the thermodynamic limit ($N \to \infty$) to account for residual two-body finite-size errors (See Supplementary Fig. 2). 

The Jastrow parameters were optimized within variational Monte Carlo using the linear optimization method of Umrigar {\it et al.}~\cite{umrigar07}.
Subsequent DMC calculations were performed with a time step of $\tau = 0.005$ Ha$^{-1}$, using size-consistent T-moves to ensure a variational treatment of the nonlocal pseudopotential terms. A vacuum region of 30~\AA~was added along the normal direction to the arsenene layers to eliminate spurious interactions between periodically replicated slabs.
Because the density operator does not commute with the Hamiltonian, we evaluated the density using the extrapolated estimator, $\rho_{\text{ext}}(r)=2\rho_{\text{DMC}}(r)-\rho_{\text{VMC}}(r)$, which cancels the leading trial-wave-function bias and leaves residual errors that are second order in the trial-wave-function error~\cite{foulkes2001quantum}.

For DFT energetics, structural relaxations, and electronic-structure calculations compared with DMC benchmarks, we performed separate DFT calculations with the VASP code~\cite{Kresse1993,Kresse1996}, solving the Kohn–Sham equations with projector-augmented wave (PAW) pseudopotentials~\cite{PAW1994,Kresse1999} for As. We used a plane-wave energy cutoff of 400 eV and an $18 \times 18 \times 1$ Monkhorst–Pack $k$-point grid for few-layer systems and an $18 \times 18 \times 4$ for bulk gray arsenic. The convergence thresholds were set to 10$^{-6}$ eV for electronic self-consistency and $5 \times 10^{-3}$ eV/\AA~for the forces during structural relaxations. We employed the semilocal LDA functional~\cite{lda}, together with several vdW-corrected approaches, namely PBE+D2~\cite{grimme2006semiempirical}, vdW-optB86b~\cite{vdW-optB88}, and SCAN+rVV10~\cite{scan_rVV10}.

\section*{Data Availability}
The data that support the findings of this study are available in
this article, its supplementary information, and the Materials
Data Facility~\cite{Blaiszik2016, Blaiszik2019} at [link to be provided upon acceptance of this manuscript].

\section*{Acknowledgements}
This work was primarily supported by the U.S. Department of Energy, Office of Science, Basic Energy Sciences, Materials Sciences and Engineering Division, as part of the Computational Materials Sciences Program and Center for Predictive Simulation of Functional Materials. 
J. Ahn (initial calculations, analysis) and J. T. Krogel (mentorship, analysis, writing) were supported by the U.S. Department of Energy, Office of Science, Basic Energy Sciences, Materials Sciences and Engineering Division, as part of the Computational Materials Sciences Program and Center for Predictive Simulation of Functional Materials.
J. Ahn also acknowledges support from U.S. Department of Energy, Office of Science, Office of Basic Energy Sciences, Computational Materials Sciences Award No. DE-SC0020177 for final calculations, analysis, and writing of the paper. 
S.-H. Kang (concept, analysis, and writing) was supported by Korea Institute of Science and Technology Information (KISTI) (K26L4M2C2-01).
An award of computer time was provided by the Innovative and Novel Computational Impact on Theory and Experiment (INCITE) program. This research used resources of the Argonne Leadership Computing Facility, which is a DOE Office of Science User Facility supported under contract DE-AC02-06CH11357. This research also used resources of the Oak Ridge Leadership Computing Facility, which is a DOE Office of Science User Facility supported under Contract DE-AC05-00OR22725.

This manuscript has been authored by UT-Battelle, LLC under Contract No. DE-AC05-00OR22725 with the U.S. Department of Energy. The United States Government retains and the publisher, by accepting the article for publication, acknowledges that the United States Government retains a non-exclusive, paid-up, irrevocable, worldwide license to publish or reproduce the published form of this manuscript, or allow others to do so, for United States Government purposes. The Department of Energy will provide public access to these results of federally sponsored research in accordance with the DOE Public Access Plan (http://energy.gov/downloads/doe-public-access-plan).

\section*{Author Contributions}
J.A. performed all calculations and analyzed the data with input from S.-H.K. and J.T.K.. J.A., S.-H.K., and J.T.K. wrote the manuscript. J.T.K. supervised the project.

\section*{Competing Interests}
The authors declare no competing interests.

\section*{References}
\bibliography{references}

@article{kim18,
  title={{QMCPACK}: an open source ab initio quantum {M}onte {C}arlo package for the electronic structure of atoms, molecules and solids},
  author={Kim, J. and Baczewski, A. D. and Beaudet, T. D. and Benali, A. and Bennett, M. C. and Berrill, M. A. and Blunt, N. S. and Borda, E. J. L. and Casula, M. and Ceperley, D. M. and others},
  journal={J. Phys.: Condens. Matter},
  volume={30},
  number={19},
  pages={195901},
  year={2018},
  publisher={IOP Publishing}
}

@article{kent20,
  title={{QMCPACK}: Advances in the development, efficiency, and application of auxiliary field and real-space variational and diffusion quantum {M}onte {C}arlo},
  author={Kent, P. R. C. and Annaberdiyev, A. and Benali, A. and Bennett, M. C. and Landinez Borda, E. J. and Doak, P. and Hao, H. and Jordan, K. D. and Krogel, J. T. and Kyl{\"a}np{\"a}{\"a}, I. and others},
  journal={J. Chem. Phys.},
  volume={152},
  number={17},
  pages={174105},
  year={2020},
  publisher={AIP Publishing LLC}
}

@article{giannozzi09,
  title={{QUANTUM ESPRESSO}: a modular and open-source software project for quantum simulations of materials},
  author={Giannozzi, P. and Baroni, S. and Bonini, N. and Calandra, M. and Car, R. and Cavazzoni, C. and Ceresoli, D. and Chiarotti, G. L. and Cococcioni, M. and Dabo, I. and others},
  journal={J. Phys.: Condens. Matter},
  volume={21},
  number={39},
  pages={395502},
  year={2009},
  publisher={IOP Publishing}
}

@article{monkhorst76,
  title={Special points for {B}rillouin-zone integrations},
  author={Monkhorst, H. J. and Pack, J. D.},
  journal={Phys. Rev. B},
  volume={13},
  number={12},
  pages={5188},
  year={1976},
  publisher={APS}
}

@article{lin01,
  title={Twist-averaged boundary conditions in continuum quantum {M}onte {C}arlo algorithms},
  author={Lin, C. and Zong, F. H. and Ceperley, D. M.},
  journal={Phys. Rev. E},
  volume={64},
  number={1},
  pages={016702},
  year={2001},
  publisher={APS}
}

@article{Kresse1993,
  title = {Ab initio molecular dynamics for liquid metals},
  author = {Kresse, G. and Hafner, J.},
  journal = {Phys. Rev. B},
  volume = {47},
  issue = {1},
  pages = {558--561},
  numpages = {0},
  year = {1993},
}

@article{Kresse1996,
  title = {Efficient iterative schemes for ab initio total-energy calculations using a plane-wave basis set},
  author = {Kresse, G. and Furthm\"uller, J.},
  journal = {Phys. Rev. B},
  volume = {54},
  issue = {16},
  pages = {11169--11186},
  numpages = {0},
  year = {1996}
}

@article{PAW1994,
  title = {Projector augmented-wave method},
  author = {Bl\"ochl, P. E.},
  journal = {Phys. Rev. B},
  volume = {50},
  issue = {24},
  pages = {17953--17979},
  numpages = {0},
  year = {1994}
}

@article{Kresse1999,
  title = {From ultrasoft pseudopotentials to the projector augmented-wave method},
  author = {Kresse, G. and Joubert, D.},
  journal = {Phys. Rev. B},
  volume = {59},
  issue = {3},
  pages = {1758--1775},
  numpages = {0},
  year = {1999}
  }

@article{scan_rVV10,
  title={Versatile van der \uppercase{w}aals density functional based on a meta-generalized gradient approximation},
  author={Peng, Haowei and Yang, Zeng-Hui and Perdew, John P and Sun, Jianwei},
  journal={Phys. Rev. X},
  volume={6},
  number={4},
  pages={041005},
  year={2016},
  publisher={APS}
}

@article{umrigar07,
  title={Alleviation of the Fermion-Sign Problem by Optimization of Many-Body Wave Functions},
  author={Umrigar, C. J. and Toulouse, J. and Filippi, C. and Sorella, S. and Hennig, R. G.},
  journal={Phys. Rev. Lett.},
  volume={98},
  number={11},
  pages={110201},
  year={2007},
  publisher={APS}
}

@article{vdW-optB88,
  title={Chemical accuracy for the van der \uppercase{w}aals density functional},
  author={Klime{\v{s}}, Ji{\v{r}}{\'\i} and Bowler, David R and Michaelides, Angelos},
  journal={J. Phys.: Condens. Matter},
  volume={22},
  number={2},
  pages={022201},
  year={2009},
  publisher={IOP Publishing}
}

@article{LDA,
  title={Self-interaction correction to density-functional approximations for many-electron systems},
  author={Perdew, John P and Zunger, Alex},
  journal={Phys. Rev. B},
  volume={23},
  number={10},
  pages={5048},
  year={1981},
  publisher={APS}
}

@article{foulkes2001quantum,
  title={Quantum Monte Carlo simulations of solids},
  author={Foulkes, William MC and Mitas, Lubos and Needs, RJ and Rajagopal, Guna},
  journal={Rev. Mod. Phys.},
  volume={73},
  number={1},
  pages={33},
  year={2001},
  publisher={APS}
}

@Article{Blaiszik2016,
author={Blaiszik, B.
and Chard, K.
and Pruyne, J.
and Ananthakrishnan, R.
and Tuecke, S.
and Foster, I.},
title={The Materials Data Facility: Data Services to Advance Materials Science Research},
journal={JOM},
year={2016},
month={Aug},
day={01},
volume={68},
number={8},
pages={2045-2052},
issn={1543-1851},
doi={10.1007/s11837-016-2001-3},
url={https://doi.org/10.1007/s11837-016-2001-3}
}

@Article{Blaiszik2019,
author={Blaiszik, Ben
and Ward, Logan
and Schwarting, Marcus
and Gaff, Jonathon
and Chard, Ryan
and Pike, Daniel
and Chard, Kyle
and Foster, Ian},
title={A data ecosystem to support machine learning in materials science},
journal={MRS Commun.},
year={2019},
month={Dec},
day={01},
volume={9},
number={4},
pages={1125-1133},
issn={2159-6867},
doi={10.1557/mrc.2019.118},
url={https://doi.org/10.1557/mrc.2019.118}
}

@article{grimme2006semiempirical,
  title={Semiempirical \uppercase{GGA}-type density functional constructed with a long-range dispersion correction},
  author={Grimme, Stefan},
  journal={J. Comput. Chem.},
  volume={27},
  number={15},
  pages={1787--1799},
  year={2006},
  publisher={Wiley Online Library}
}

@article{zhu2015strain,
  title={Strain-induced metal-semiconductor transition in monolayers and bilayers of gray arsenic: A computational study},
  author={Zhu, Zhen and Guan, Jie and Tom{\'a}nek, David},
  journal={Phys. Rev. B},
  volume={91},
  number={16},
  pages={161404},
  year={2015},
  publisher={APS}
}

@article{kecik2016stability,
  title={Stability of single-layer and multilayer arsenene and their mechanical and electronic properties},
  author={Kecik, D and Durgun, Engin and Ciraci, S},
  journal={Phys. Rev. B},
  volume={94},
  number={20},
  pages={205409},
  year={2016},
  publisher={APS}
}

@article{ersan2019two,
  title={Two-dimensional pnictogens: A review of recent progresses and future research directions},
  author={Ersan, Fatih and Ke{\c{c}}ik, Deniz and {\"O}z{\c{c}}elik, VO and Kadioglu, Y and Akt{\"u}rk, O {\"U}zengi and Durgun, Engin and Akt{\"u}rk, Ethem and Ciraci, S},
  journal={Appl. Phys. Rev.},
  volume={6},
  number={2},
  pages={021308},
  year={2019},
  publisher={AIP Publishing}
}

@article{kamal2015arsenene,
  title={Arsenene: Two-dimensional buckled and puckered honeycomb arsenic systems},
  author={Kamal, C and Ezawa, Motohiko},
  journal={Phys. Rev. B},
  volume={91},
  number={8},
  pages={085423},
  year={2015},
  publisher={APS}
}

@article{shah2020experimental,
  title={Experimental evidence of monolayer arsenene: an exotic 2\uppercase{D} semiconducting material},
  author={Shah, Jalil and Wang, Weimin and Sohail, Hafiz Muhammad and Uhrberg, RIG},
  journal={2D Mater.},
  volume={7},
  number={2},
  pages={025013},
  year={2020},
  publisher={IOP Publishing}
}

@article{zhao2017magnetotransport,
  title={Magnetotransport properties in a compensated semimetal gray arsenic},
  author={Zhao, Lingxiao and Xu, Qiunan and Wang, Xinmin and He, Junbao and Li, Jing and Yang, Huaixin and Long, Yujia and Chen, Dong and Liang, Hui and Li, Chunhong and others},
  journal={Phys. Rev. B},
  volume={95},
  number={11},
  pages={115119},
  year={2017},
  publisher={APS}
}

@article{bullett1975density,
  title={Density of states calculation for crystalline \uppercase{A}s and \uppercase{S}b},
  author={Bullett, DW},
  journal={Solid State Commun.},
  volume={17},
  number={8},
  pages={965--967},
  year={1975},
  publisher={Elsevier}
}

@article{xu1993tight,
  title={Tight-binding theory of the electronic structures for rhombohedral semimetals},
  author={Xu, JH and Wang, EG and Ting, CS and Su, WP},
  journal={Phys. Rev. B},
  volume={48},
  number={23},
  pages={17271},
  year={1993},
  publisher={APS}
}

@article{gonze1990first,
  title={First-principles study of \uppercase{A}s, \uppercase{S}b, and \uppercase{B}i electronic properties},
  author={Gonze, Xavier and Michenaud, J-P and Vigneron, J-P},
  journal={Phys. Rev. B},
  volume={41},
  number={17},
  pages={11827},
  year={1990},
  publisher={APS}
}

@article{arcudia2020blue,
  title={Blue phosphorene bilayer is a two-dimensional metal--and an unambiguous classification scheme for buckled hexagonal bilayers},
  author={Arcudia, Jessica and Kempt, Roman and Cifuentes-Quintal, Miguel Eduardo and Merino, Gabriel and Heine, Thomas},
  journal={Phys. Rev. Lett.},
  volume={125},
  pages={196401},
  year={2020},
  publisher={APS},
}

@article{kadioglu2018diffusion,
  title={Diffusion quantum Monte Carlo and density functional calculations of the structural stability of bilayer arsenene},
  author={Kadioglu, Yelda and Santana, Juan A and {\"O}zaydin, H Duygu and Ersan, Fatih and Akt{\"u}rk, O {\"U}zengi and Akt{\"u}rk, Ethem and Reboredo, Fernando A},
  journal={J. Chem. Phys.},
  volume={148},
  pages={214706},
  year={2018},
  publisher={AIP Publishing}
}

@article{wang2019new,
  title={A new generation of effective core potentials from correlated calculations: 4s and 4p main group elements and first row additions},
  author={Wang, Guangming and Annaberdiyev, Abdulgani and Melton, Cody A and Bennett, M Chandler and Shulenburger, Luke and Mitas, Lubos},
  journal={J. Chem. Phys.},
  volume={151},
  number={14},
  pages={144110},
  year={2019},
  publisher={AIP Publishing}
}

@article{schiferl1969crystal,
  title={The crystal structure of arsenic at 4.2, 78 and 299 \uppercase{K}},
  author={Schiferl, D and Barrett, CS},
  journal={J. Appl. Crystallogr.},
  volume={2},
  number={1},
  pages={30--36},
  year={1969},
  publisher={International Union of Crystallography}
}
\bibliographystyle{naturemag}
\end{document}